# Suppressing Metal–Molecule Charge Transfer with a Phosphorus Interlayer

Mattia Bassotti,*,a,b Luca Floreano,c Luca Schio,c Sergio Salaverria,d Dimas G. de Oteyza,d,e Ggiacomo Giorgi,f,g,h,i Frederik Schiller j and Alberto Verdini.*,b

*a* Department of Physics, University of Perugia, 06123 Perugia, Italy

*b* CNR-IOM – Istituto Officina dei Materiali, Department of Physics, 06123 Perugia, Italy

*c* CNR-IOM – Istituto Officina dei Materiali, Area Science Park, 34149 Trieste, Italy

*d* Nanomaterials and Nanotechnology Research Center (CINN), CSIC-UNIOVI-PA, 33940 El Entrego, Spain

*e* Donostia International Physics Center, 20018 San Sebastián, Spain

*f* Department of Civil & Environmental Engineering (DICA), University of Perugia, 06125 Perugia, Italy

*g* CNR-SCITEC, 06123 Perugia, Italy

*h* CIRIAF - Interuniversity Research Centre, University of Perugia, 06125 Perugia, Italy

*i* Centro S3, CNR-Istituto Nanoscienze, 41125 Modena, Italy

*j* Centro de Fisica de Materiales, 20018 San Sebastián, Spain

## Abstract

Porphyrins are organic molecules that exhibit excellent opto-electronics properties, making them suitable for a variety of applications. Nevertheless, their functionality strongly depends on the surface onto which they are deposited, and on the interaction between the molecules and the substrate itself, which often leads to an undesired alteration in their electronic properties. In this study, we use a phosphorus interlayer on a Cu(110) surface as a buffer layer for the electronic decoupling of Zinc-TetraPhenylPorphyrin (ZnTPP) molecules. Using a combination of complementary techniques, such as Near Edge X-ray Absorption Fine Structure (NEXAFS), X-ray and Ultraviolet Photoemission Spectroscopy (XPS, UPS) as well as Scanning Tunneling Spectroscopy (STS) techniques, it is shown how the charge transfer from the metal, responsible for quenching the ZnTPP lowest unoccupied molecular level (LUMO) levels, is effectively prevented by the presence of a phosphorus reconstruction in between.

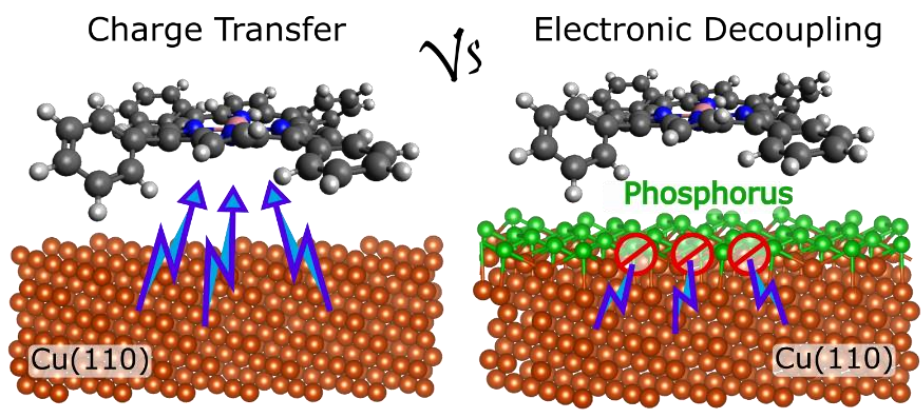

**TOC Abstract:** Phosphorus forms stable reconstructions with remarkable properties on various metal substrates. Here, using a spectroscopic approach, we demonstrate its role as an effective buffer layer on Cu(110), preventing charge transfer to adsorbed Zn-tetraphenylporphyrins (ZnTPP). This interlayer ensures electronic decoupling at the metal–molecule interface, preserving the intrinsic electronic structure of the molecular layer.

## Introduction

Porphyrins are versatile organic molecules with key roles in organic electronics [1], chemical sensing [2], and catalysis [3], due to their high thermal stability, chemical tunability, and ability to form stable, ordered monolayers [4,5]. However, the integration of organic materials into functional devices (for instance Organic Field Effect Transistors, Organic Light-Emitting Diodes and Organic Photovoltaics) generally involves interfaces to semiconductors and/or metals, where the strong substrate–molecule interactions can alter their geometry and electronic structure [6–9], even leading to transmetalation processes in some cases [10,11]. Minimizing such interactions is thus crucial for preserving the intrinsic molecular functionality.

Among the coinage metals, commonly used as contact electrodes, the copper surfaces are the most reactive ones, where the contact organic layer is typically metallized by charge transfer, whether not chemically bound. The Cu surface can be effectively passivated by intercalation of an insulating monolayer of NaCl before molecular deposition [12,13]; as an alternative, predeposition of sub-monolayers amount of Sn may suitably tune the Cu reactivity to organic small molecules [14–16]. An alternative route involves the growth of an oxide interlayer before molecule deposition, which suppresses the charge transfer from the metal and preserves the molecules electronic features [17,18].

Nevertheless, these materials are not without limitations: they often require specific growth conditions, are restricted to a narrow set of substrates, and in some cases lack sufficient thermal stability. While these approaches are encouraging, the search for new strategies remains a priority, particularly those that can be generalized across a wide range of systems and substrates, providing scalable solution for implementing organic material in devices.

In this context, phosphorus emerges as a promising candidate. Stable 2D phosphorus layers have been successfully grown by molecular beam epitaxy (MBE) on several metal surfaces, including Cu [19,20], Pt [21,22], Au [23,24], Ag [25,26] and Ni [27]. On these substrates, phosphorus can form a variety of ordered structures and phases, while many other have been theoretically predicted [28], making it a versatile platform for templating the growth of organic molecules with different shapes and sizes.

Beyond its structural versatility, phosphorus offers unique electronic properties, including a strain-tunable direct band gap and high hole mobility, which have contributed to its growing interest as a novel 2D material [29–32]. These features make phosphorus an ideal candidate not only as a decoupling interlayer, but also as an electronic and structural template for organic–inorganic heterostructures, combining molecular functionalities with the tunability of 2D phosphorus.

Despite its promise, only one study to date has explored phosphorus–organic molecule heterostructures, demonstrating how a blue phosphorene monolayer enhances molecular ordering on Au(111) [33]. This highlights the potential of phosphorus in this context and pushes the investigation into its role as a decoupling layer.

In this work, we investigate the well-studied Zinc(II) meso-tetraphenylporphyrin (ZnTPP) as a model system to test the effectiveness of a phosphorus buffer layer for electronic decoupling. Its stability and well-characterized photophysical properties make it valuable for studying fundamental processes like electron transfer and energy transfer, as well as for designing functional materials and devices. We employed red phosphorus—a cost-effective alternative to black phosphorus—as the precursor to grow a P monolayer on Cu(110), which served as a template for subsequent ZnTPP deposition. We combined Near-Edge X-ray Absorption Fine Structure (NEXAFS), X-ray Photoemission Spectroscopy (XPS), Scanning Tunneling Spectroscopy (STS), Ultraviolet Photoemission Spectroscopy (UPS), and Density Functional Theory (DFT) to study the resulting system.

Our results demonstrate that the phosphorus interlayer preserves the electronic structure of the ZnTPP. Direct adsorption of ZnTPP on bare Cu(110) leads to strong hybridization and an alignment of the molecular orbitals that brings the Lowest Unoccupied Molecular Level (LUMO) below the Fermi level, with a consequently important charge transfer. In contrast, the phosphorus layer effectively reduces the molecule-metal interaction, realigns the molecular orbitals and suppresses the charge transfer. These findings underscore the potential of phosphorus-based interlayers for the scalable integration of porphyrins into electronic and optoelectronic devices.

**Results and discussion**

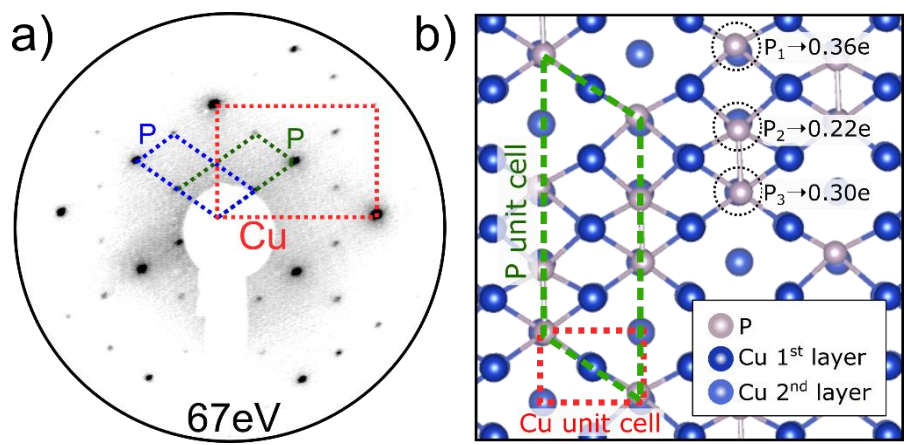

*Figure 1. a) LEED image (67 eV energy) of P on Cu(110). The substrate Cu(110) unit cell is traced in red, while in green and blue are the unit cells corresponding to the two rotational domains of the P reconstruction. b) Model of the P/Cu(110) system. The unit cell is drawn in green for P and red for Cu. The P atoms (grey) forming the reconstruction are non-equivalent and originate three different local electrical charges.*

We grow the phosphorus film on Cu(110) by following the procedure described by Zhang et al. [20], except for replacing the black phosphorus with the red phosphorus as a precursor, in line with the work of Sala et al. [34] The resulting P superstructure is a $\sqrt{3} \times \sqrt{}(11/2)$ – R55° reconstruction with two rotational domains, visible in the LEED image of Figure 1a, where the P and substrate unit cells are highlighted showing the same structure found by Zhang et al. [20] . Further confirmation of the equivalence between red and black P as precursors comes from the P 2p, Cu 3s and valence band photoelectron spectra (Figure S1) following the two-step deposition procedure (Figure S1). Both LEED and XPS results confirm red phosphorus as a suitable and cost-effective substitute for black phosphorus in MBE growth of monolayer P on Cu(110).

To further characterize the P/Cu(110) system, which will later serve as a template for ZnTPP monolayer growth, we carried out DFT calculations. The atomic model shown in Figure 1b displays both Cu(110) and P unit cells, consistent with literature reports [20].

The model reveals that P atoms preferentially occupy hollow sites on the Cu(110) surface, resulting in three chemically distinct P species. This is reflected in the calculated local charge densities, which are 0.30e, 0.22e, and 0.36e, respectively (with e=-1.6·10$^{-19}$ C being the elementary charge).

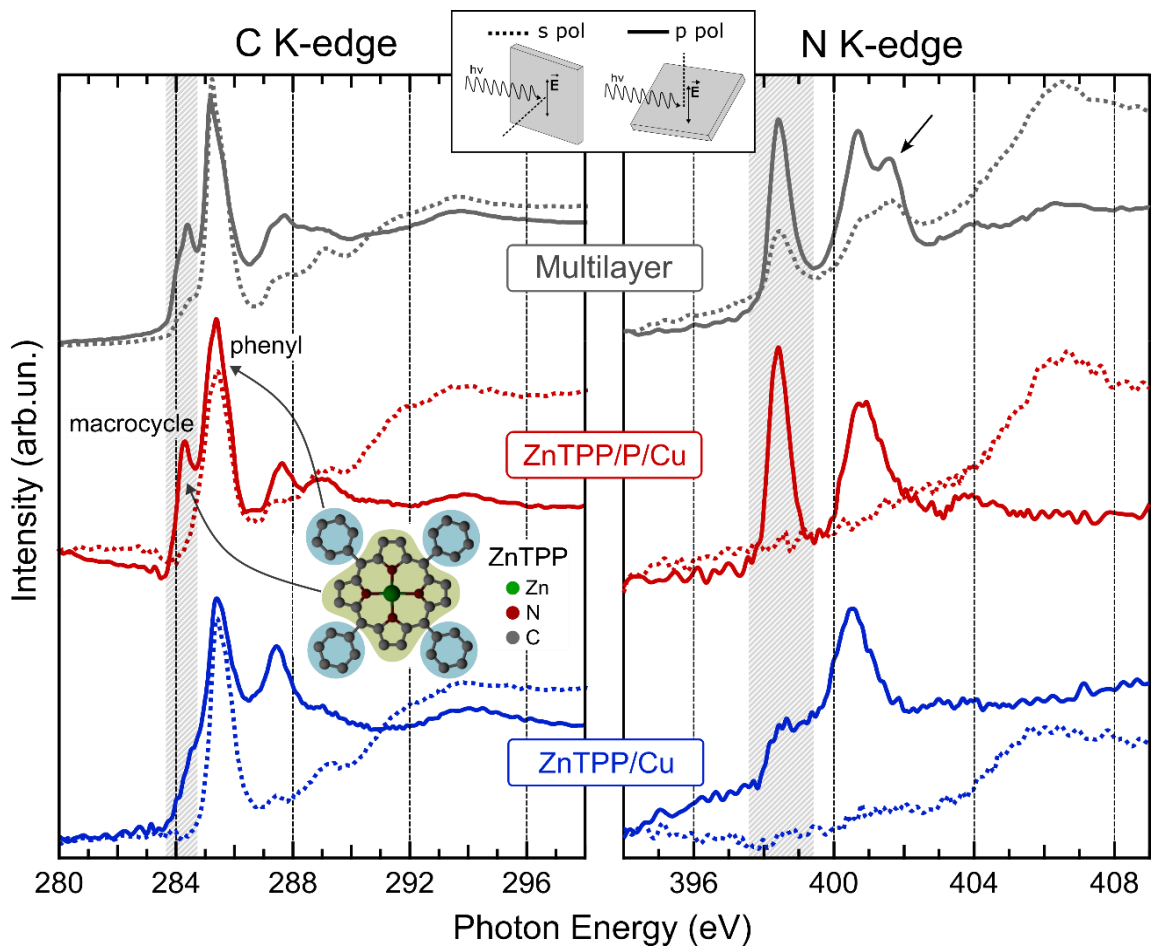

*Figure 2. C K-edge and N K-edge NEXAFS spectra for the ZnTPP/Cu, ZnTPP/P/Cu and multilayer systems, from bottom to top, respectively. Each polarization is individually addressed: in close to p polarization geometry (solid line) the electric field is closely parallel (6°) to the surface normal, while in the s polarization geometry (dashed line) the photon polarization is parallel to the surface, as visible in the inset. The grey areas highlight the LUMO levels for which the role of the P buffer layer can be appreciated more, preventing the corresponding absorption peak to be quenched by the electron transfer from the substrate. The arrow in the N K-edge graph points to the high-order unoccupied molecular orbital resonance that is quenched in the monolayer systems. The ZnTPP molecule is shown for visualizing the macrocycle and phenyl contributions in the C K-edge spectra.*

NEXAFS measurements provide evidence of the electronic configuration of the unoccupied molecular orbitals, as well as the adsorption geometry of the organic molecular layer. In order to visualize the effectiveness of a P buffer layer, we compare the C and N K-edge NEXAFS spectra for a monolayer (ML) ZnTPP growth on P/Cu(110) with a ML ZnTPP growth on the bare Cu(110) and a multilayer of ZnTPP, as shown in Figure 2. The monolayer directly grown on Cu(110) is the benchmark for evaluating the intensity of the interaction between the metallic substrate and the organic molecule. The multilayer system, instead, provides a reference for the electronic properties of (almost) non-interacting molecules [35,36]. Being the NEXAFS spectra recorded in Partial Electron Yield (PEY) mode, in which the detected secondary electrons emitted from the sample are produced within a limited escaped depth, the sensitivity is limited to the very top layers, enough distanced from the surface to not experience any significant effects. Moreover, intermolecular interactions within the multilayer are assumed to be negligible [5].

For each system, the NEXAFS spectra have been collected using two distinct sample orientation with respect to the photon beam polarization vector, namely "p" and "s" polarization, with a constant grazing angle of 6°. As illustrated in the inset of Figure 2, in the s (p) polarization the incident photons have the polarization vector parallel (perpendicular) to the sample surface [37]. Since the photon absorption process depends on the relative orientation between the involved molecular orbitals and the photon polarization, the NEXAFS dichroism can be used to deduce the average molecular orientation with respect to the substrate surface. This is particularly effective for studying a π-conjugated molecule, like ZnTPP [5]. The results suggest flat lying molecules in all cases considered here (see below).

The detailed assignment of the absorption features lies beyond the scope of this work, and it is discussed elsewhere [38]. However, for the purposes of the following discussion, it is important to note that the N 1s and C 1s spectra highlight different regions of the molecule and thus offer complementary information [38]. In particular, the N 1s excitation includes only moderate contributions from the phenyl moieties. Similarly, the spatial distribution of the molecule gives rise to two main C 1s excitation contributions [38], associated with the phenyl groups and the macrocyclic ring, as illustrated in the inset of Figure 2.

Looking first at the C K-edge, the feature at lower photon energy, highlighted by the grey shadow area in the graph, is commonly associated to the 1s → π* transition related to the C atoms of the macrocycle, while the feature at higher photon energy (295eV) corresponds to the same type of transition but for the C atoms in the phenyl groups (see the ZnTPP model inside the graph) [35,38–40]. For all the three systems we observe a strong linear dichroism for the LUMO state associated with the porphyrin macrocycle, in which the intensity in p-polarization is significantly larger than the corresponding one in s-polarization, indicating a flat absorption geometry of the molecule atop the surface for both the P buffer layer and bare Cu(110) surfaces. In contrast, the phenyl-related features exhibit a noticeable contribution from s-polarization, suggesting that the phenyl moieties adopt a tilted orientation relative to the substrate, in agreement with previously reported adsorption geometries for monolayer porphyrins on surfaces [36,39].

Regarding the intensity of the LUMO state, a comparison across the three systems clearly shows a different behavior. While the phenyl-related feature is almost identical in all cases, the macrocycle-related one is well-visible only in the multilayer and the P buffered systems. This feature is significantly suppressed when considering the ZnTPP/Cu case. The same scenario is observed in the N K-edge spectra, where the 1s → π* absorption feature at about 398.4 eV [35,38–40], also marked by a grey shaded area and likewise associated with the macrocycle, is diminished in the ZnTPP/Cu case.

The suppression of an absorption feature indicates an electronic charge transfer from the metallic substrate to the molecule due to a strong coupling between them, the electrons from the substrate populate the molecule's formerly unoccupied molecular orbitals, rendering those states unavailable for absorption transitions. Such behavior has already been reported for porphyrin on reactive surfaces [41–43]. However, in the presence of a phosphorus buffer layer, this charge transfer is effectively inhibited, particularly at the macrocycle, since the corresponding absorption peak remains visible, indicating that the transition is still allowed.

It is worth noting that higher-energy unoccupied molecular orbital features, such as the one indicated by the arrow in the N K-edge spectrum of the ZnTPP multilayer, are partially quenched in both the ZnTPP/P/Cu and ZnTPP/Cu systems. This suggests that these higher unoccupied states are somehow affected to some degree by the P reconstruction. Nonetheless, the phosphorus interlayer effectively prevents charge transfer from the metallic substrate, as evidenced by the overall preservation of the main absorption features in the ZnTPP/P/Cu system.

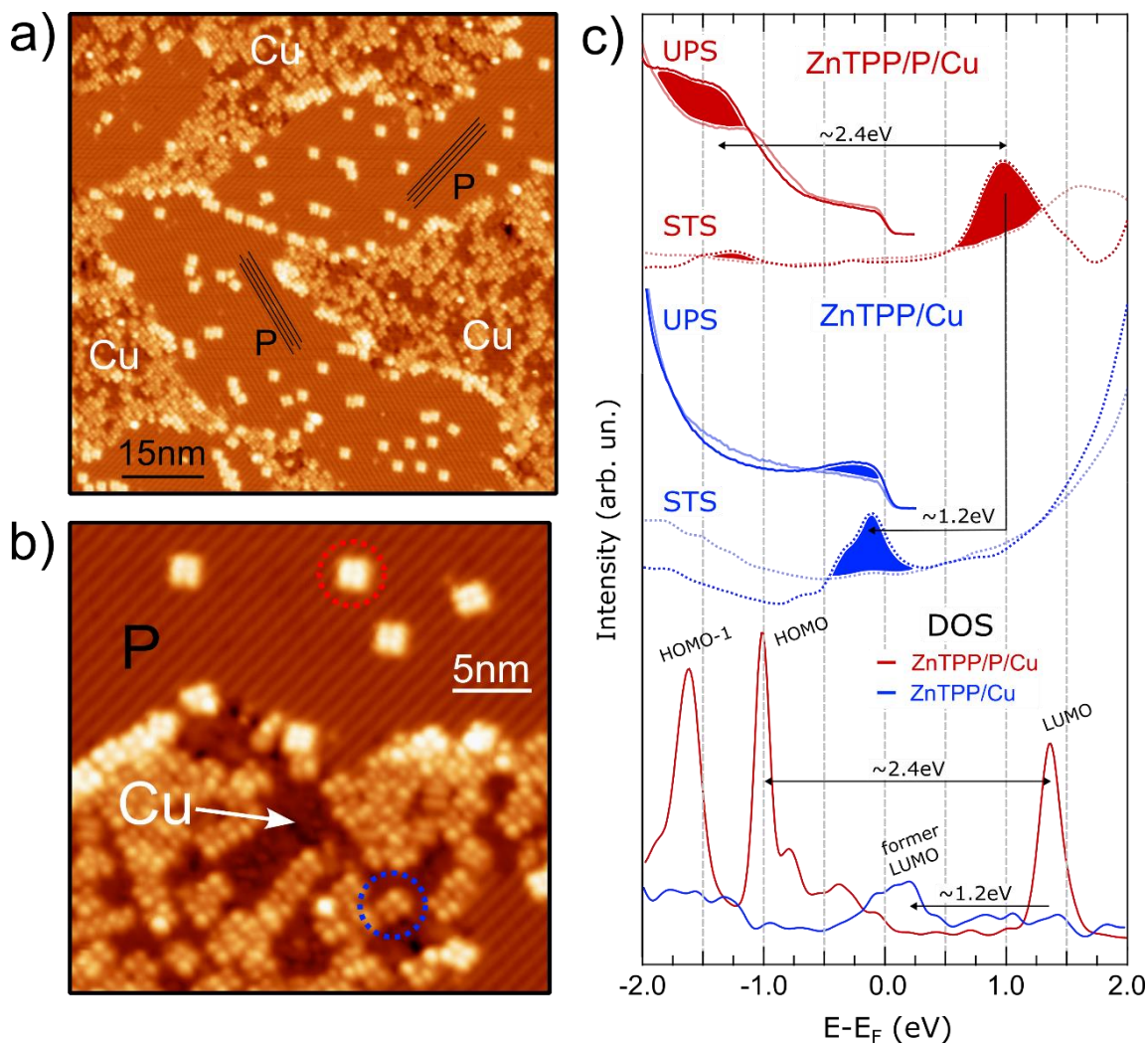

*Figure 3. a) Wide-range STM image showing a low-coverage ZnTPP deposition on the system formed by the two P domains partially covering the Cu(110) substrate. Image size = 90 nm², U = 1.25 V, I= 80pA. b) Close-up STM image showing the molecules involved for the STS: on top of P (red) and on top of bare Cu (blue). Image size = 30 nm², U = 1 V, I= 80pA. c) UPS (solid lines) and STS (dotted lines) spectra for the ZnTPP/P/Cu system (in blue) and the ZnTPP/Cu system (in red). Reference spectra on the respective substrates are shown in lighter colours. The main features on the spectra are highlighted by filled colour areas. On the bottom, the local density of states, as calculated by DFT simulations.*

Ultraviolet Photoemission Spectroscopy (UPS) and Scanning Tunneling Spectroscopy (STS) further corroborate and deepen the insights obtained from NEXAFS measurements.

For STM and STS measurements, a low coverage of ZnTPP was deposited on an incomplete film of P on Cu(110). Figures 3a and 3b show, respectively, large-scale and high-resolution STM images of the system used for this purpose. It should be noticed that this system was not obtained through direct deposition of phosphorus onto the substrate. Instead, we utilized P atoms originating from earlier depositions, which had diffused into the bulk during annealing cycles at 500°C [44,45]. These atoms subsequently migrated back to the surface upon low-temperature annealing at 300°C for 10 minutes, giving rise to the coexistence of the P stripes and clean Cu phases.

The P reconstruction visible in the STM images is consistent with those reported in the literature [20] and reported in Figure 1, with two rotational domains that, in this case, only partially cover the substrate, leaving exposed regions of bare Cu(110) and allowing the study of Cu and P-anchored molecules within the very same system, an ideal scenario for direct comparative measurements with the same STM tip.

A first notable observation from the STM images is that ZnTPP molecules adsorb preferentially on the bare Cu(110) areas, further supporting the stronger interaction between the molecules and the metal substrate compared to the phosphorus-covered regions. This also means that the molecules are able to travel at room temperature on the P and Cu(110) regions.

Molecules highlighted by red and blue circles in Figure 3b are representative to those used for acquiring the STS spectra shown in Figure 3c, corresponding to ZnTPP/P/Cu and ZnTPP/Cu systems, respectively. The STS spectra are displayed alongside UPS spectra, allowing not only a cross-validation of the conduction band measurements but also a reliable comparison between the low-coverage system used for STS and the full monolayer coverage investigated with UPS, NEXAFS, and XPS (see discussion below). For comparison, STS and UPS spectra of the underlying substrate without ZnTPP is added to better visualize the new spectroscopic features.

Upon deposition of ZnTPP on the bare Cu surface, the UPS spectrum clearly shows an increase of intensity just below the Fermi level (blue curves), which may be associated with the partially filled LUMO, from comparison with the NEXAFS spectra. When the molecule is deposited on the P/Cu(110) interface instead, no change is observed in the Fermi region, but a new state is recorded at -1.55 ± 0.05 eV. This state closely matches the binding energy value of 1.65-1.75 eV reported in the literature for the Highest Occupied Molecular Level (HOMO) level associated with the macrocycle of ZnTPP in a multilayer [46,47]. The small shift with respect to the multilayer is also identical to that reported for one monolayer of ZnTPP on the passivated Fe(001)-$p$(1×1)O surface, as due to the interface dipole [47].

Focusing on the ZnTPP/P/Cu system, STS clearly resolves two distinct resonances within the probed energy range, namely around -1.4 eV and 1 eV. We associate these to the HOMO and LUMO levels, respectively. Instead, only one clearly resolved resonance is observed on ZnTPP/Cu, centred around -0.2 eV but with its high energy tail clearly crossing the Fermi level. We associate this resonance to the LUMO, whose energy level alignment denotes a downward shift of ~1.2 eV with respect to the decoupled molecules on P, resulting in its notable but not yet complete filling by charge transfer [48]. A similar behavior has been found in the case of organic molecules on fcc(110) surfaces, such as Ag(110) [40]. The same shift brings the HOMO energy out of the probed energy range and therefore not visible in our measurements.

This picture is in excellent agreement with the NEXAFS measurements. Whereas the LUMO level remains empty and available for π* transitions for ZnTPP/P/Cu, it gets almost entirely filled for the molecules directly on Cu, which consequently quenches the associated NEXAFS transitions.

At the bottom of Figure 3c, we report the results of Density Functional Theory (DFT) calculations of the local density of states (LDOS), showing the sum of the element and orbital contributions of the molecule (N, C, H, and Zn; s and p orbitals) with or without the P layer.

The calculations are based on the structural models shown in Figure S2, where ZnTPP is adsorbed either on top of the phosphorus layer or directly on Cu(110). For each configuration, several anchoring positions were considered, and only the most energetically stable geometry was used for the LDOS analysis (see Supporting Information for details).

The relative energy positions of the features calculated at the DFT level of theory show good agreement with those obtained from STS and UPS data. In particular, the LDOS for ZnTPP/P/Cu exhibits a sharp peak in the conduction band, approximately 2.4 eV above the HOMO, which is absent when ZnTPP is adsorbed directly on Cu(110). In the ZnTPP/Cu LDOS, the former LUMO level appears significantly broadened and shifted by ~1.2 eV towards lower energy, now centered close to the Fermi level.

Some underestimation in the theoretical data is expected. This arises from several factors, including: (i) the ground-state nature of DFT, which inherently struggles to describe excited-state properties, (ii) the use of Zinc-Tetramethylporphyrin instead of Zinc-Tetraphenylporphyrin in the calculations to reduce computational cost (see SI), and (iii) the neglect of coverage effects (e.g., molecular anchoring density) in the anchoring mechanism model. Despite these limitations, the agreement between the experimental and theoretical relative peak positions is highly encouraging.

Overall, the results further highlight the partial filling of the LUMO due to electron injection from the metallic substrate, illustrating the extent of charge transfer and corroborating the experimental spectroscopic findings.

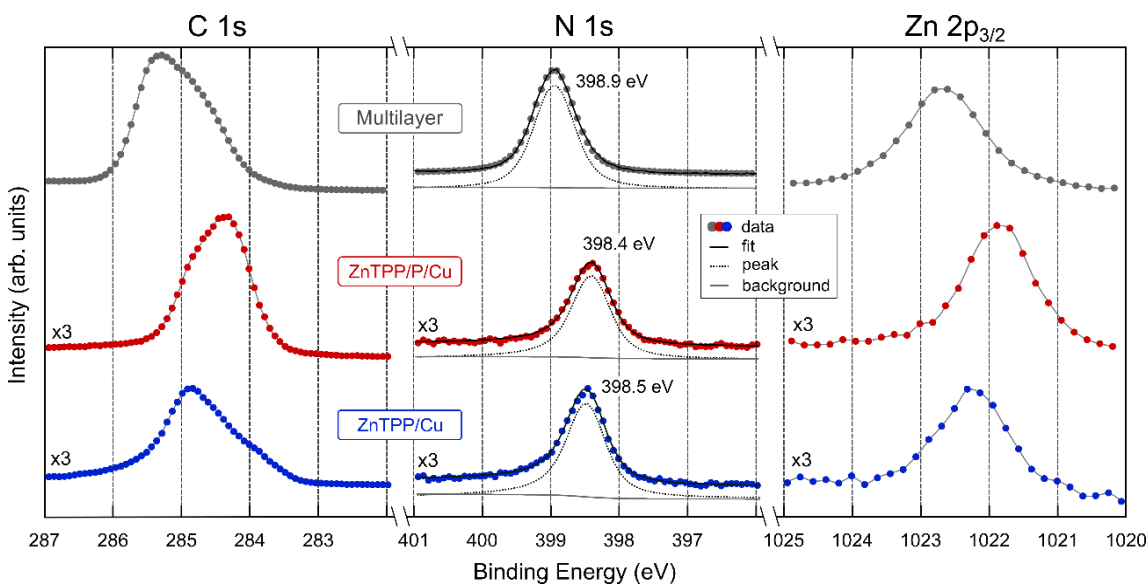

*Figure 4. C 1s, N 1s and Zn 2p$_{3/2}$ XPS spectra for a ML of ZnTPP on Cu and a ML of ZnTPP on Cu with a P buffer layer, and a multilayer from bottom to top, respectively. In the first two cases the intensity has been multiplied by a factor 3 for the sake of clarity. The N 1s spectra are shown with their best fit envelopes, while the C 1s and Zn 2p$_{3/2}$ data points are shown with eye-guide lines, in grey.*

Finally, figure 4 presents the C 1s, N 1s, and Zn 2p3/2 XPS spectra for the multilayer, ZnTPP/Cu, and ZnTPP/P/Cu systems (top to bottom, respectively). There is a shift of approx. 0.5 eV towards lower binding energy (BE) going from the multilayer to the monolayer on clean Cu(110). This trend is expected as also already found in similar works [8].

An additional shift towards lower binding energies with respect to the ZnTPP/Cu system is observed for the porphyrins adsorbed on the phosphorus layer. The shifts in the XPS peaks of the ZnTPP/P/Cu system relative to the ZnTPP multilayer can be attributed to the screening effect induced by the P/Cu substrate. Notably, the shifts differ across atomic species, and in the case of the C 1s core level, variations are also observed in the peak shape.

It is important to highlight that the C 1s signal in the free molecule arises from the superposition of four distinct contributions, corresponding to four non-equivalent carbon atoms in the porphyrin structure [39]. Upon adsorption on the phosphorus layer, this peak broadens and changes shape, likely due to the influence of the reconstructed P layer, which contains three non-equivalent phosphorus atoms with different local charge densities. As a result, the local chemical environment experienced by each ZnTPP atom varies depending on its adsorption site, leading to a redistribution of the binding energies in the XPS spectra.

## Conclusion

This work investigates the role of phosphorus as a buffer layer for electronically decoupling a monolayer of the organic molecule ZnTPP from a metallic substrate, specifically Cu(110). Three systems were studied and compared: (i) a multilayer of ZnTPP, (ii) a monolayer of ZnTPP deposited directly on Cu(110), and (iii) a monolayer of ZnTPP grown on the reconstructed phosphorus layer on Cu(110).

NEXAFS analysis reveals that, in the presence of the phosphorus buffer layer, the LUMO levels of ZnTPP are largely preserved, exhibiting spectral features comparable to those of the multilayer reference. In contrast, when ZnTPP is directly adsorbed on Cu(110), these unoccupied states are quenched due to charge transfer from the metallic substrate. This interpretation is supported by STS and UPS measurements, as well as DFT calculations, which show a distinct feature in the valence band of the ZnTPP/Cu system corresponding to the now-filled LUMO level, confirming the occurrence of electron transfer from the substrate.

Furthermore, XPS measurements of the C 1s, N 1s, and Zn $2p_{3/2}$ core levels exhibit binding energy shifts in both ZnTPP/P/Cu and ZnTPP/Cu systems relative to the multilayer. These shifts can be attributed to variations in the local environment experienced by ZnTPP molecules, depending on their adsorption site, particularly on the reconstructed phosphorus layer composed of three non-equivalent P atoms with different local charge densities.

Overall, the results strongly support the use of phosphorus as an effective buffer layer for the growth of well-ordered organic monolayers on metal substrates, significantly mitigating the charge transfer that typically compromises the intrinsic electronic properties of organic molecules. This approach becomes even more attractive considering the successful use of red phosphorus, a cost-effective alternative to black phosphorus, as a precursor in the growth process.

## Experimental Details

All the sample preparations and measurements were performed in Ultra-High Vacuum (UHV) conditions (with a pressure better than $10^{-7}$ Pa). The Cu(110) sample was cleaned by repeated cycles of sputtering (Ar+, 1 keV, 10 mA) and annealing (500 °C) and the surface quality and order checked through Low-Energy Electron Diffraction (LEED) or Reflection High-Energy Electron Diffraction (RHEED). If not differently specified, red phosphorus (99.99% purity, Merck) was evaporated from a Knudsen cell and sublimated on the cleaned Cu(110) kept at RT. To get the P reconstruction, the sample has been subsequently annealed to 320 °C for 30 minutes. ZnTPP molecules (99.99% purity, Merck) were sublimated from a crucible at 300 °C, after evaporation rate control through a quartz microbalance, with the sample kept at RT.

XPS and NEXAFS measurements took place at the Elettra synchrotron (Trieste, Italy), ALOISA beamline.

The XPS core level measurements were performed at normal emission and grazing incidence with photon energies of 1180 eV for Zn 2p, 495 eV for N and C 1s, P2p, Cu3s, Cu3p. All the XPS spectra were calibrated relative to the measured Fermi level.

NEXFAS measurements of the N and C K-edges were performed in partial electron yield (PEY) mode, through an electrostatic high-pass filter used to cut off secondary electrons (-370 eV and -250 eV for N and C K-edges, respectively). The spectra were taken in two distinct geometries: keeping the light incidence angle at 6° and rotating the sample around the beam axis to have the surface either parallel (s-polarization) or perpendicular (p-polarization) to the photon electric field. For the absorption spectra normalization, clean samples were measured and the drain current on the last refocusing mirror was acquired, to take into account photon flux and beam fluctuations.

UPS data were acquired at a laboratory-based setup (CFM, San Sebastián, Spain) using Helium Iα excitation line (hν = 21.2 eV). A hemispherical analyzer (SPECS-150) was used with angular and energy resolution set to 0.1° and 40 meV, respectively. The shown spectra are taken at 45° emission angle and integrating the channelplate (+/-15°) perpendicular to the slit.

The STM and STS data were acquired on a commercial low-temperature scanning tunnelling microscope (University of Oviedo, Mieres, Spain) with a chemically etched tungsten tip. The measurement temperature was 78 K and the spectroscopy was acquired with an integrated lock-in applying an oscillation amplitude of 20 mV.

**Computational Details**

DFT calculations have been performed by means of the VASP code [49–52].The Generalized Gradient Approximation (GGA) as implemented by Perdew, Burke, and Ernzerhof (PBE) has been exploited to solve the exchange-correlation part of the potential [53]. The projector-augmented wave (PAW) potentials [54], with an energy cutoff of 500 eV, have been adopted. The D3(BJ) dispersion correction contribution has been included in all the calculations [55]. Geometries were considered converged when forces were lower than 0.05 eV/Å.

We start considering the cubic polymorph of copper (space group 225; Fm3m, 4 atoms) whose optimization led to $a=b=c=3.56$ Å, in good agreement with experimental values [56]. On top of the unit cell, we at first obtained the Cu(110) orientation and later we considered and optimized a large supercell (8×8×4, 448 Cu atoms) with $a=20.15$ and $b=28.49$ Å, sufficiently large to accommodate the porphyrin system. Due to the large size of the system, we optimized the structure just sampling the Brillouin Zone at the Γ point.

Keeping fixed the Cu ions of the bottom layer of the slab, we added a large amount of vacuum on top of the surface to prevent any possible spurious interactions among replicas along the non-periodic direction. We considered several anchoring of the porphyrin system on top of the P covered/P-free Cu(110) surface.

The charge analysis is performed by means of the Bader code [57–60]. The formation energy of the several adducts considered in this work formed by porphyrin and Cu(110) surface (with and without P atoms) is calculated as:

$$E_{form} = E_{add} - [E_{porphyrin} + E_{Cu(110)}]$$

where $E_{add}$ is the total energy of the final adduct, while $E_{porphyrin}$ and $E_{Cu(110)}$ are the energy of the porphyrin and of the Cu (110) slab, respectively.

**ACKNOWLEDGMENT**

A.V. and M.B. gratefully acknowledge Stefano Cristiani for his valuable assistance.

**Corresponding Authors**

Mattia Bassotti bassotti@iom.cnr.it

Alberto Verdini verdini@iom.cnr.it

**Present Addresses**

**Author Contributions**

Mattia Bassotti: methodology, investigation, data curation, visualization, and writing – original draft. Luca Schio: investigation, reviewing. Luca Floreano: supervising, investigation, reviewing. S. Salaverria: investigation, reviewing. D. G. de Oteyza: supervising, investigation, reviewing. Giacomo Giorgi: simulations, reviewing, data visualization. Frederik Shiller: supervising, investigation, reviewing. Alberto Verdini: conceptualization, methodology, investigation, writing – original draft, reviewing, funding acquisition, project administration, and supervision.

**Key Words**

Porphyrin, Decoupling, Interlayer, Red-Phosphorus, NEXAFS

**Funding Sources**

This work has been partially funded by the European Union - NextGenerationEU, Mission 4, Component 2, under the Italian Ministry of University and Research (MUR) National Innovation Ecosystem grant ECS00000041 - VITALITY - CUP B43C22000470005 and by the project MEGS, PRIN 2022 "MultiExciton Generation in tailored molecular heterojunctionS (MEGS)" – grant 20224PJT7C CUP B53D23003930006.

This work was also supported by grant PID2022-140845OB-C64 and PID2023-149158OB-C44, funded by the Spanish MICIU/AEI/10.13039/501100011033 and FEDER Una manera de hacer Europa. Additional support from the Basque Government grant IT-1591-22 is acknowledged.

**Notes**